\documentclass[10pt]{article}
\pdfoutput=1
\usepackage{amsmath}
\usepackage{amssymb}

\usepackage{graphicx}

\usepackage{cite}

\usepackage{color} 

\usepackage[labelfont=bf,labelsep=period,justification=raggedright]{caption}

\makeatletter
\renewcommand{\@biblabel}[1]{\quad#1.}
\makeatother

\date{}

\newcommand{\panel}[1]{\textbf{(\uppercase{#1)}}}
\begin{document}

\begin{flushleft}
{\Large
\textbf{Self-sustained activity, bursts, and variability in recurrent networks}
}
\\
Marc-Oliver Gewaltig$^{1,2}$
\\
1) Blue Brain Project, \`Ecole Polytechnique F\`ederale de Lausanne, 
QIJ, CH-1015 Lausanne, Switzerland\\
2) Honda Research Institute Europe GmbH, D-63073 Offenbach, Germany\\
E-mail: marc-oliver.gewaltig@epfl.ch
\end{flushleft}

\section*{Abstract}
There is consensus in the current literature that stable states of
asynchronous irregular spiking activity require (i) large networks of
$10\,000$ or more neurons and (ii) external background activity or
pacemaker neurons.
Yet already in 1963, Griffith showed that networks of simple threshold
elements can be persistently active at intermediate rates.  Here, we
extend Griffith's work and demonstrate that sparse networks of
integrate-and-fire neurons assume stable states of self-sustained
asynchronous and irregular firing without external input or pacemaker
neurons. These states can be robustly induced by a brief pulse to a
small fraction of the neurons, or by short a period of irregular
input, and last for several minutes.
Self-sustained activity states emerge when a small fraction of the
synapses is strong enough to significantly influence the firing
probability of a neuron, consistent with the recently proposed
long-tailed distribution of synaptic weights. 
During self-sustained activity, each neuron exhibits highly irregular
firing patterns, similar to experimentally observed
activity. Moreover, the interspike interval distribution reveals that
neurons switch between discrete states of high and low firing rates.
We find that self-sustained activity states can exist even in small
networks of only a thousand neurons. We investigated networks up to
$100\,000$ neurons.
Finally, we discuss the implications of self-sustained activity for
learning, memory and signal propagation.

\section*{Author Summary}
Neurons in many brain areas are active even in the absence of a
sensory stimulus. Many models have tried to explain this
\emph{spontaneous activity} by spiking activity, reverberating in
recurrent networks of excitatory and inhibitory neurons. But so far
the conclusions have been that such networks can only sustain
spontaneous activity under certain conditions: The networks must be
large and there must be either endogeneously firing neurons (so called
\emph{pacemaker neurons}) or diffuse external input which keeps the
network active.  Here we show that recurrent networks of excitatory
and inhibitory neurons can sustain spontaneous activity for periods of
many minutes, provided that a small percentage of the connections are
sufficiently strong. Thus, contrary to previous findings,
self-sustained (spontaneous) activity does neither require large
networks nor external input or pacemaker neurons. The spike patterns
observed during self-sustained activity are chaotic and highly
irregular. The interspike interval distribution during self-sustained
activity reveals that the network switches between different discrete
states, each characterized by their own time scale.  Our results
provide a possible explanation of self-sustained cortical activity and
the role of the recently observed long-tailed weight distributions in
the mammalian cortex.

\section*{Introduction}
Spontaneous activity, that is, activity in the absense of a sensory
stimulus, is a ubiquitous phenomenon in the brain that has puzzled
generations of researchers. 
Spontaneous activity is highly irregular and has a strong effect on
evoked neuronal responses \cite{Arieli1996,Shadlen1998}. In fact many
researchers argue that this \emph{ongoing activity} represents
information rather than noise\cite{Tsodyks99,Fukushima2012}. Moreover,
spontaneous activity in the cortex is stable and robust and can be
observed in awake as well as in anesthesized animals throughout their
entire life.

It is commonly assumed that spontaneous activity is created by
reverberating activity within recurrent neuronal circuits, but the
exact mechanisms by which neuron maintain their low rate firing are
still not well understood.

Already in 1956 Beurle showed that networks of excitatory neurons (``a
mass of units capable of emitting regenerative pulses'') generally
have an inherently unstable activity in which all or none of the units
are excited \cite{Beurle1956}.

In 1962 Ashby and co-workers \cite{Ashby1962} reproduced Beurle's
findings under more simplified assumptions and delivered a
mathematical proof for the instability of recurrently connected
excitatory threshold units. They derived an expression for the
probability of an output pulse as function of the probability for an
input pulse and showed that this function takes the form of the now
well known sigmoid. They concluded that ``the more richly
organized regions of the brain offer us something of a paradox. They
use threshold intensively, but usually transmit impulses at some
moderate frequency, seldom passing in physiological conditions into
total inactivity or maximal excitation. Evidently there must exist
factors or mechanisms for stability which do not rely on fixed
threshold alone'' \cite{Ashby1962}.

In 1963 Griffith extended the models of Beurle \cite{Beurle1956} and
Ashby \cite{Ashby1962} in two ways. First, he remarked that
networks with dedicated connectivity can support stable states of low
or intermediate firing rates. For example the \emph{complete transmission
  line} consists of consecutive groups of neurons that are connected
by diverging/converging connections. In such a network, activity will
travel unperturbed from one group to the next, without exciting the
entire network. This network architecture became later known as the
\emph{synfire chain} \cite{Abeles1982}. 

The second addition of Griffith were inhibitory neurons which Ashby
had neglected. Griffith also found that networks of excitatory and
inhibitory neurons don't support stable states at low or
intermediate levels of activity if the neurons have many excitatory
and inhibitory inputs with correspondingly small synaptic weights.

He then went on to show that for the special case of a few synaptic
inputs per neuron, such stable activity states should indeed
exist. Since computational power in 1963 was more limited than it is
today, he restricted his analysis to the case of few excitatory inputs
with global inhibition, i.e. and one inhibitory input, strong enough to
suppress the combined input of all excitatory neurons. In this case,
activity is stable at an intermediate rate. This suggests that stable,
low or intermediate firing rates should exist for more realistic
network configurations.

Thirty years later, van Vreeswijk and Sompolinski revived interest in
self-sustaining activity in recurrent networks with two seminal papers
\cite{Vreeswijk1996,Vreeswijk1998} in which the authors introduced the
concept of \emph{balanced excitation and inhibition} as a criterion
for the emergence of stable activity states. But in their model,
external input is needed to obtain stable activity at low or
intermediate rates.

Since then several studies have shown that self-sustained activity is
possible under some conditions.  In recurrent networks of
\emph{conductance based} neurons, self-sustained activity can survive
for a limited time\cite{Kuhn2004,Kumar2008}. Otherwise additional
activating mechanisms, like external input or
\cite{Amit1997,Brunel2000}, endogeneously firing
neurons\cite{Latham2004a,Vogels2005}, or cells which respond to an
\emph{inhibitory} stimulus\cite{Destexhe2008}, are needed to sustain
activity.

Moreover, self-sustained activity in recurrent network models is still too
regular compared to experimental data\cite{Softky1993a,Barbieri2008}
and additional \emph{de-correlating} mechanims are needed\cite{Barbieri2007}.

In this paper we show that highly irregular self-sustained activity is
an inherent property of recurrent neural networks with excitation and
inhibition. These states occur in relatively small networks (one
thousand neurons and more) if the connectivity is sparse and the
connection strengths is large.

Self-sustained activity can be robustly induced by a brief pulse to a
small fraction of the neurons. We will show that these self-sustained
states differ in their survival time statistics and their interspike
interval distributions from previously reported self-sustained states.

In the following section, we will briefly revisit the results of Ashby
\cite{Ashby1962} and Griffith \cite{Griffith1963} to show under which
conditions recurrent networks of excitatory and inhibitory neurons can
sustain states of low rate almost indefinitely. We will then
investigate the nature and properties of these states in computer
simulations. We demonstrate that self-sustained activity, even in
small networks is stable and long-lived, provided the connectivity is
sparse and synapses are strong. We compare self-sustained activity
with the asynchronous irregular state (AI state), characterized by
Brunel \cite{Brunel2000}. We find that the firing patterns of neurons
in the self-sustained states are highly irregular. This irregularity
can be seen in the wide range of firing rates and large coefficients
of variation (CV) of the interspike intervals. While self-sustained
activity states require sparse and strong connections, the required
post-synaptic potential (PSP) amplitudes are still in the
physiological range. Moreover, self-sustained activity states emerge
in weakly coupled networks with a few strong connections,
corresponding to the recently proposed idea that cortical connectivity
is best described as a few strong synapses in a sea of weak ones
\cite{Song2005} or a long-tailed distribution of synaptic weights.

\section*{Results}
\subsection*{Stability in networks of simple threshold elements}

Following Ashby \cite{Ashby1962}, we describe the activity of a neuron
by a single number $p$ which satisfies $0 \leq p \leq 1$ and which is
the probability of observing a spike in a sufficiently small interval
$\Delta t$:
\begin{gather}
p= \lim_{t\rightarrow \infty} \left( \frac{n}{C}\cdot\frac{\Delta t}{t}\right)
\end{gather}
where $n$ is the number of spikes arriving at the inputs $C$ at time
$t$.  The probability of observing exactly $n$ spikes on the $C$
inputs follows a binomial distribution:
\begin{gather}
  p_n = {C \choose n }p^n(1-p)^{C-n}
\end{gather}

Now assume that a neuron fires, if there are $n \geq \theta$ input
spikes. Then the probability for producing an output spike is given by
the cumulative binomial probability function:
\begin{gather}
  P(p)=\sum_{n=\theta}^C {C \choose n }p^n(1-p)^{C-n} \label{eq:ashby1}
\end{gather}

We can estimate the long-term behavior of the network activity by
considering equation \eqref{eq:ashby1} as an iterative map. Starting
from an initial activity $p_0$, we repeatedly apply
\eqref{eq:ashby1}. Ashby found, that the only stable fixed points in
this iteration are $p=0$ and $p=1$, that is, either all the neurons
are silent or all neurons are active.  Griffith \cite{Griffith1963}
noted that this situation changes if the network also contains
inhibitory neurons. In the case of $C_E$ excitatory and $C_I$
inhibitory inputs, the new threshold condition becomes:
\begin{gather}
n_E-g \cdot n_I \ge \theta \label{eq:griffith1}
\end{gather}
where $n_E$ is the number of spikes at the $C_E$ excitatory inputs and
$n_I$ the number of spikes at the $C_I$ inhibitory inputs, and $g$ a
factor that captures the difference in synaptic efficacies. To reflect
the cortical ratio of 80\% excitatory neurons to 20\% inhibitory
neurons, we choose:
\[
C_I=\gamma \cdot C_E
\]
with $\gamma=0.25$.

To obtain the probability for passing the threshold, we must now
consider the joint probabilities for observing $n_E$ excitatory and
$n_I$ inhibitory spikes. The respective cumulative probability
function given by:
\begin{gather}
  P(p)= \sum_{n_E-g\cdot n_E \geq \theta} {C_E \choose n_E}{C_I
    \choose n_I} p^{n_E+n_I}q^{C_E+C_I-n_E-n_I} \label{eq:griffith2}
\end{gather}
where $q=1-p$ and the sum is over all combinations of $n_E$ and $n_I$
that satisfy the threshold condition \eqref{eq:griffith1}.
Again, we can use this expression as an iterative map to determine
whether a given probability $p$ is stable under repeated application
of equation \eqref{eq:griffith2}, that is:
\begin{gather}
p^*=p=P(p)
\end{gather}
with the condition that
\begin{gather}
\left|\left(\frac{\partial P(p)}{\partial p}\right)_{p^*}\right| < 1 \label{eq:stability}
\end{gather}

Griffith showed that no stable activity exists, if $C_E$ and $C_I$
become large. But, if the number of inputs is small and the
inhibitory efficacy is very strong, then there is a stable solution at
$p=1/2$. Griffith solved equation \eqref{eq:griffith2} for the case of
global inhibition which is strong enough to suppress an output spike,
even if all excitatory inputs are active.

Unfortunately, it is not easy to determine whether other stable
solutions to equation \eqref{eq:griffith2} exist, because the equation
is discrete and involves the binomial coefficients of all possible
input combinations that reach or exceed threshold. 
Typical approximations use the assumptions that the number of inputs
is large and the individual probabilities are low. Then, the law of
large numbers allows us to replace the binomial distribution with a
normal distribution. Unfortunately, these are the conditions for which
we already know that all solutions are unstable \cite{Griffith1963}
and that external input is needed to obtain self-reproducing
activities \cite{Amit1997,Vreeswijk1998,Brunel2000}.

In this paper we won't attempt to simplify equation \eqref{eq:griffith2}
further, but rather we will investigate it numerically. We will demonstrate
that for sparse networks with large synaptic efficacies, there indeed
exist stable self-sustaining activity states. We will then investigate
these states in large-scale network simulations of current based
integrate-and-fire neurons.
In a companion paper by Enger et al. (2013) we derive a theory which
explains self-sustained activity states as well as their survival
times.

\subsubsection*{Many inputs and high-threshold}

\begin{figure}[th]
\includegraphics{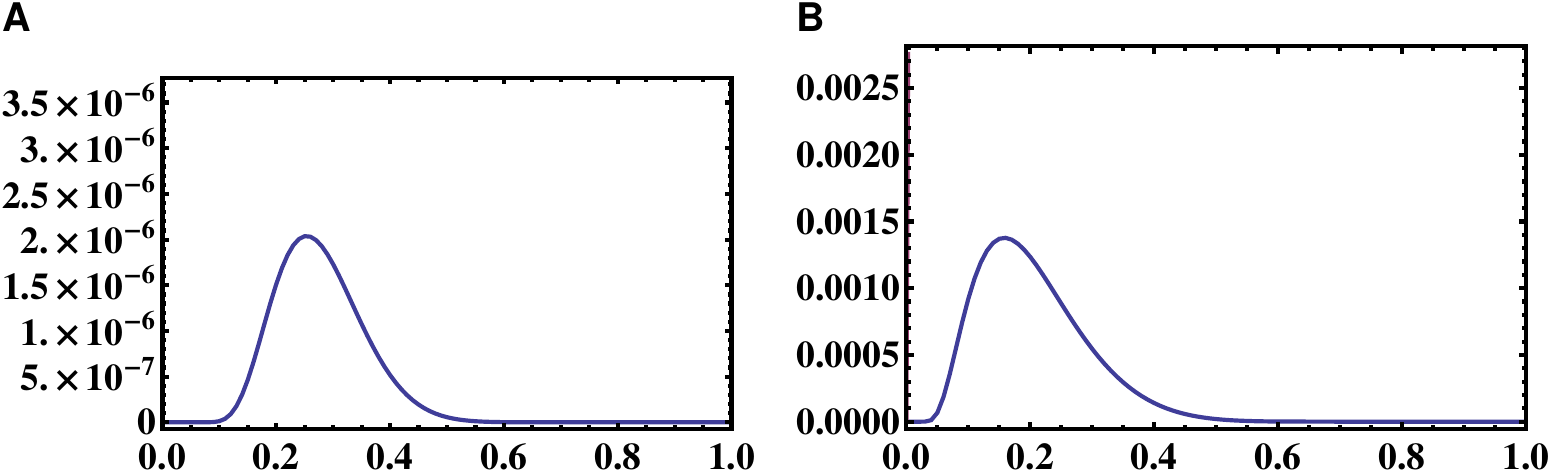}
\caption{\label{fig:1} Effect of the firing threshold on the firing
  probability transfer function (numerical solution of
  equation \eqref{eq:griffith2}) \panel{a} Many synapses, high threshold; $
  \theta=100, C_E=1\,000, C_I=250$. The probability stays orders of
  magnitudes below the diagonal and the only stable fixed-point is
  $P(0.0)=0.0$. \panel{b} Many synapses, lower threshold; $\theta=50,
  C_E=1\,000, C_I=250$. The peak probability is increased by almost
  three orders of magnitude.}
\end{figure}

Neurons in cortex have a large number of synapses
\cite{Braitenberg1991,Abeles1991} and it is often assumed that this
also implies that the individual synapses must be small
\cite{Abeles1991}, that is, the threshold is large. Unfortunately, it
is difficult to reach threshold under these conditions, because a
large number of neurons must be simultaneously active.
This is reflected in the very low output probability for this
situation. Figure \ref{fig:1}A shows the numerical solution of
equation \eqref{eq:griffith2} for a network with $C_E=1\,000, C_I=250,
\theta=100$. The probability stays orders of magnitudes below the
diagonal and the only stable fixed-point is $P(0.0)=0.0$.

Incidentally, these parameters are used by Brunel \cite{Brunel2000}
for the asynchronous irregular activity state. Since the output
probability is low, the asynchronous irregular state requires an
external input that lifts the output probability above the diagonal.

The only way to increase the firing probability without adding
external input is to lower the firing threshold.  In figure
\ref{fig:1}B, the threshold is only half as high at $\theta=50$ and as
a result the peak probability has increased by a factor of almost
$1\,000$, but the curve is still far below the diagonal.

\subsubsection*{Stability with low threshold}

The threshold has a strong non-linear influence on the firing
probability, because the odds of finding sufficiently many coincident
(or near coincident) spikes decrease at least exponentially with
increasing threshold.  We can use this lever to our advantage and lift
the firing probability above the diagonal and thus generating stable
self-sustaining activity. We do this by considerably lowering the
firing threshold.

\begin{figure}[th]
\includegraphics{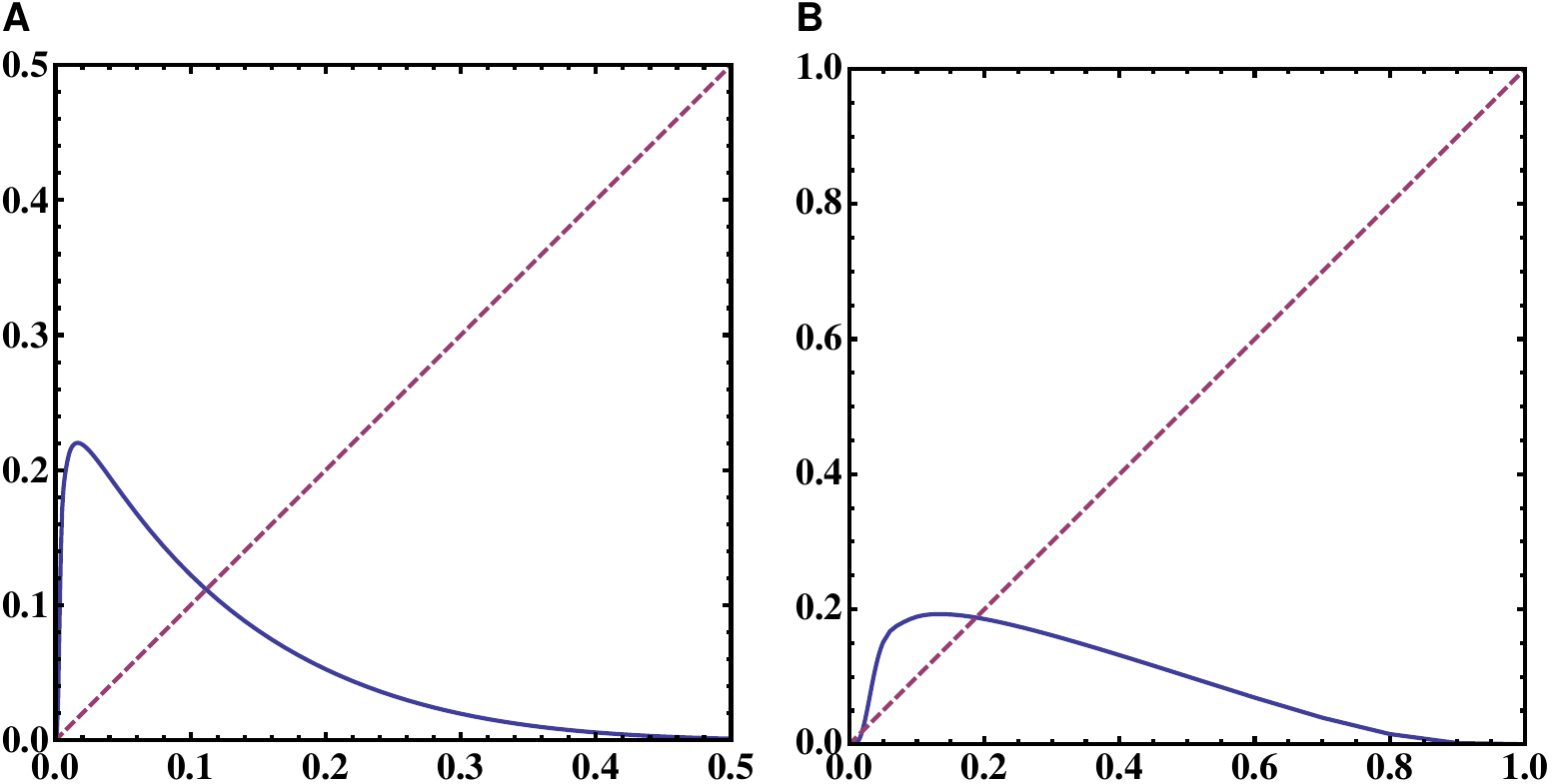}
\caption{\label{fig:2} Stable points of the firing probability
  transfer function (numerical solution of equation
  \eqref{eq:griffith2}) \panel{a} Many synapses, low threshold; $
  \theta=5, C_E=1\,000, C_I=250$. The probability steeply rises to its
  peak at $p=0.004$, before is falls towards zero again. The
  intersection with the diagonal at $p^* \approx 0.11$ as a stable
  fixed point. \panel{b} Low threshold and fewer synapses ($C_E=100,
  C_I=25, \theta=5$). The curve rises more slowly than in (a) and also
  falls off more slowly towards zero. The intersection with the
  diagonal at $p^* \approx 0.2$ is a stable fixed point. }
\end{figure}

Figure \ref{fig:2}A shows the probability for $\theta=5$ ($C_E=1\,000,
C_I=250$). This time, the probability crosses the diagonal, indicating
that there are stable, self-reproducing activity states.  The curve
has three important points $p^+, p*$, and $p^-$.

First, the \emph{ignition point} $p^+$ , defined as
\begin{equation} 
 p^+= \min \{p \in (0,1] \text{ with } P(p) = p \}. \label{eq:ignition-point}
\end{equation}

Second, the \emph{stable self-reproducing} point $p^*$, defined as
\begin{equation} 
 p^*= \min \{p \in (p^+,1] \text{ with } P(p) = p \}.
\end{equation}

Third, the \emph{shut-off} point $p^-$, beyond which all activity ceases
again. It is defined as: 
\begin{equation} 
 p^-= \min \{ p \in (p^*,1] \text{ with } P(p) = p^+\}.
\end{equation}

For $p<p^+$ the firing probability quickly tends to zero, for $p^+ <
p < p^*$ the firing probability is amplified, and for $p>p^*$ the
probability $P(p)$ quickly tends to zero again. Thus, stable activity
is only possible within the range $p^+ < p < p^-$. If too few or too many
neurons in the network are active, activity will cease in the next
iteration.

In figure \ref{fig:2}A we have $p^+ \in (0.0006,0.0007)$, $p^* \approx
0.11$, and $p^-\in(0.5,0.6)$. $P(p)$ quickly rises towards its maximum
at $p=0.004$, where the probability is amplified by a factor of 34.
It then quickly falls towards zero. Since the activity in the network
will move along this curve, we expect the rates in such a network to
be volatile. If the activity exceeds the shut-off point $p^-=0.6$,
activity will cease in the next iteration.

Figure \ref{fig:2}B demonstrates that the stable point persists, even
if we reduce the number of synapses by a factor of 10 ($C_E=100,
C_I=25, \theta=5$). The probability still rises above the diagonal and
crosses it again at $p^*\approx 0.2$. Decreasing the number of
connections has obviously increased the activity at the stable point,
because with $g=5>1/\gamma$, the synaptic weights decrease more
strongly than the number of connections.

Another notable change is the slope of the curve. The probability
rises more gently, has a wider peak, and also decreases more slowly,
compared to \ref{fig:2}A. Thus, we expect the rates in a network with these
parameters to be less volatile. Moreover, the shut-off point $p^-$ is
beyond $0.9$, thus, the network can endure much higher activities
without shutting off.

\subsection*{Stability in recurrent neural networks}

We now turn from numerical evaluations of the simplified Griffith
model to simulations of recurrent networks of current based
integrate-and-fire neurons. In these simulations we investigate
whether the stable states of self-sustaining activity found in
Griffith's model can indeed be induced in networks of spiking neurons.

\subsubsection*{Neuron and network model}
The network model that we will be using is based on the sparse random
network by Brunel \cite{Brunel2000}.  It consists of $N_{E}$
excitatory and $N_{I}$ inhibitory neurons, with $N_{I}=\gamma
N_{E}$ and $\gamma=0.25$.

Each neuron receives input from $C_{E}$ excitatory and $C_{I}$
inhibitory neurons, with $C_{E}=\epsilon N_{E}$ and $C_{I}=\epsilon
N_{I}$, where $\epsilon$ satisfies $0\leq \epsilon \leq 1$. The cases
$\epsilon=0$ and $\epsilon=1$ correspond to an unconnected and a fully
connected network, respectively.

We consider integrate-and-fire neurons with current based synapses
\cite{Tuckwell1}, whose membrane potential is given by:
\begin{equation}
\tau_m\frac{d V_{m}^i}{d t}(t)=-V_{m}^i(t)+R_mI^{i}(t),\label{eq:potential}
\end{equation}
for each neuron $i=1\ldots N=N_{E}+N_{I}$, where $\tau_m$ is the membrane
time constant, $R_m$ the membrane resistance, and $I^i(t)$ the
synaptic current.  Whenever the membrane potential $V_m$ reaches the
threshold value $V_{th}$, a spike is send to all post-synaptic
neurons.

Each spike induces a post-synaptic current, modeled as alpha
functions:
\begin{gather}
  psc(t) = \alpha \cdot t\exp\left(-\frac{t}{\tau_{syn}} \right),
\end{gather}
where $\alpha$ is chosen such that the resulting post-synaptic
potential has amplitudes $J_E$ and $J_I$ for excitatory and inhibitory
synapses, respectively.

$J_{ij}$ is the efficacy and $D_{ij}$ the delay of the synapse from
neuron $j$ to neuron $i$. Excitatory synapses have efficacy $J_{E}$
and inhibitory synapses efficacy $J_{I}=-g\cdot J_{E}$, with $g>0$.
The parameters $g$ and $\gamma$ determine the ratio of excitation to
inhibition. The regime $g\approx1/\gamma$ is called the
\emph{balanced} regime, $g>1/\gamma$ the inhibition dominated regime,
and $g<1/\gamma$ the excitation dominated regime.

The details of our model are summarized in the figures
\ref{fig:netparams} and \ref{fig:neuronparams}.

\subsubsection*{Relation to Griffith's model}

The parameters of Griffith's model are the number of excitatory and
inhibitory inputs $C_E$ and $C_I$, the ratio between excitation
and inhibition $g$, and the threshold $\theta$.

$C_E$ and $C_I$ are determined by the respective
number of excitatory and inhibitory neurons, $N_E$ and $N_I$, as well
as the connection probability $\epsilon$.

The threshold $\theta$ is given by the membrane threshold $V_{\theta}$ and
the excitatory synaptic weight $J_E$:
\begin{gather}
  \theta = V_{\theta}/J_E
\end{gather}

Before we turn to stable states of self-sustained activity, we will
review the model Brunel \cite{Brunel2000} which describes how external
excitatory spike input can induce low-rate activity in a recurrent
network. We use this model as a reference for comparison with the
self-sustained states that don't require external input.

Next, we consider the cases of stable self-sustained activity,
discussed in the previous section. We will then extend our simulations
and investigate how the firing rate and the survival time of
self-sustained states depend on the ratio of excitation and inhibition
as well as the size of the excitatory synaptic weight $J_E$.

\subsubsection*{Brunel's model: many connections, high threshold}

In his model of asynchronous irregular activity, Brunel
\cite{Brunel2000} assumed that each neuron has a large number of
synapses with accordingly small synaptic weights. In Griffith' terms,
these assumptions correspond to the case of many connections with high
threshold and we have seen that in this regime, there is no stable
activity state except for $p=0$. To overcome this problem the model is
usually supplied with external excitatory input
\cite{Amit1997,Brunel2000} or pacemaker neurons
\cite{Latham2000,Vogels2005}.

\begin{figure}[htp]
\includegraphics{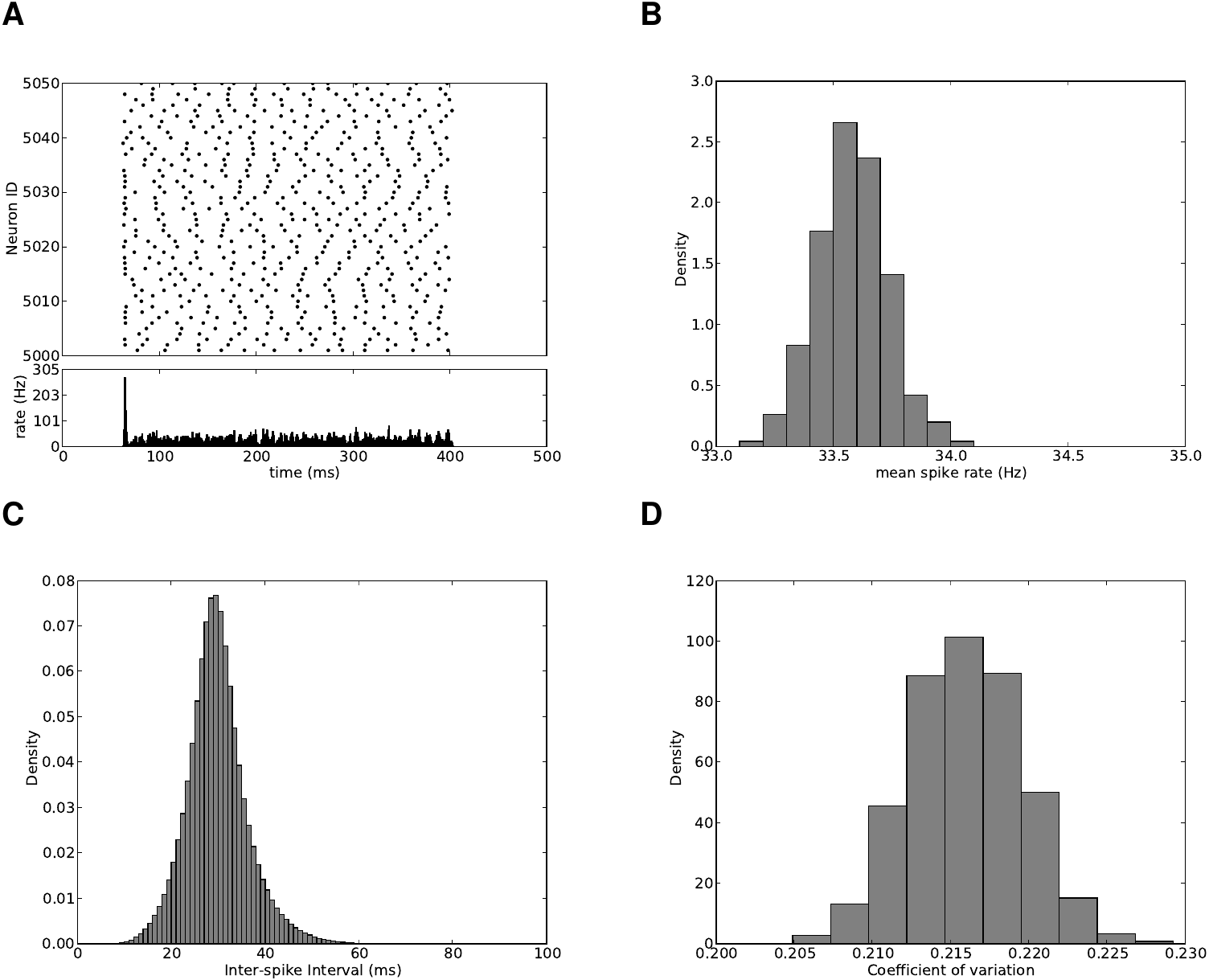}
\caption{\label{fig:brunel} Activity in a random network with
  $12\;500$ neurons, $g=5$, $J=0.1$, $\epsilon=0.1$, and Poisson
  background activity with $\nu_{ext}=38\;$Hz. Panels B-D are computed
  from simulations that lasted 100 seconds. \panel{a} Raster plot
  (top) and spike count (bottom) of 50 neurons for 500
  milliseconds. Poissonian background activity is supplied between
  $t=50\,ms$ and $t=400\,ms$. \panel{b} Firing rate distribution of
  all excitatory neurons. The firing rates are approximately
  Gaussian distributed within the range of $33\;$Hz to $34.5\;$Hz.  \panel{c}
  interspike interval distribution.  The interspike intervals are
  Gaussian distributed with a mean of $30\;ms$ and standard deviation
  of approx. $6\;ms$. \panel{d} Histogram of the coefficient of
  variation. The coefficients of variation of the interspike intervals
  are also Gaussian distributed with mean $0.22$ and standard deviation
  $0.004$. }
\end{figure}

Figure \ref{fig:brunel} shows the spiking activity, induced by an
external Poisson input, as proposed by Brunel \cite{Brunel2000}.  The
configuration corresponds to the case, used in figure
\ref{fig:1}A and to the \emph{asynchronous irregular state}, shown in
figure 8C of \cite{Brunel2000}.

Figure \ref{fig:brunel}A shows a raster plot of the spiking
activity of 50 neurons over an epoch of $500\;ms$. Each point in the
raster-plot corresponds to a spike of a neuron at the respective
time. At time $t=0\;ms$ all neurons are in the quiescent state. At
$t=50\;ms$ an external Poisson input is switched on and induces
spiking activity in the network. The Poisson input then persists for
the duration of the simulation. After an initial transient the
network assumes the asynchronous irregular state (AI)
\cite{Brunel2000}. This state is not self-sustaining, because if the
Poisson input is switched off at $t=400\;ms$ the network falls back
into the quiescent state again.

Barbieri and Brunel \cite{Barbieri2008} observed that the spiking
activity produced by these types of networks cannot explain the
irregularity of experimentally recorded data. This is illustrated in
figures \ref{fig:brunel}B-\ref{fig:brunel}D. 

Figure \ref{fig:brunel}B shows that the firing rates in the network,
measured over $100$ seconds follow a very narrow Gaussian distribution
with mean $33.5\;$Hz and standard deviation less than $1\;$Hz. In
other words, all neurons fire essentially at the same rate. The small
variation of the rates argues against the assumption that each neuron
is firing according to a Poisson distribution. Figure
\ref{fig:brunel}C confirms this by showing that also the interspike
intervals follow a narrow Gaussian distribution ($30\;ms \pm 6\;ms$) ,
as opposed to an exponential distribution that would be the signature
of Poissonian firing.  The coefficients of variation that result from
this narrow ISI distribution (Figure \ref{fig:brunel}D) are
accordingly small (CV $\approx 0.215 \pm 0.05$). Thus, the neurons in
this network behave like oscillators whose period is slightly
perturbed.

\subsection*{Self-sustained activity in recurrent networks of integrate-and-fire
  neurons}

Griffith's model predicts that stable self-sustained activity states
exist in networks where the firing threshold is sufficiently low. We
will first look for these states in simulation. Next we will
investigate, how abundant and how robust these self-sustained activity
states are. Do they exist only for a very limited set of parameters,
or do they exist in a larger region of the parameter space? We will
then have a closer look at the firing pattern of self-sustained
activity.

Our starting point is the configuration shown in figure
\ref{fig:2}B. Assuming a firing threshold of $V_{th}=20\;mV$,
Griffith's model predicts that we can observe stable self-sustained
activity with an excitatory synaptic weight of
$J_E=20\;mV/5=4\;mV$. Indeed, in simulation we found that a network
with $J_E=4\;mV$ shows self-sustained activity states.

\begin{figure}[htp]
\includegraphics{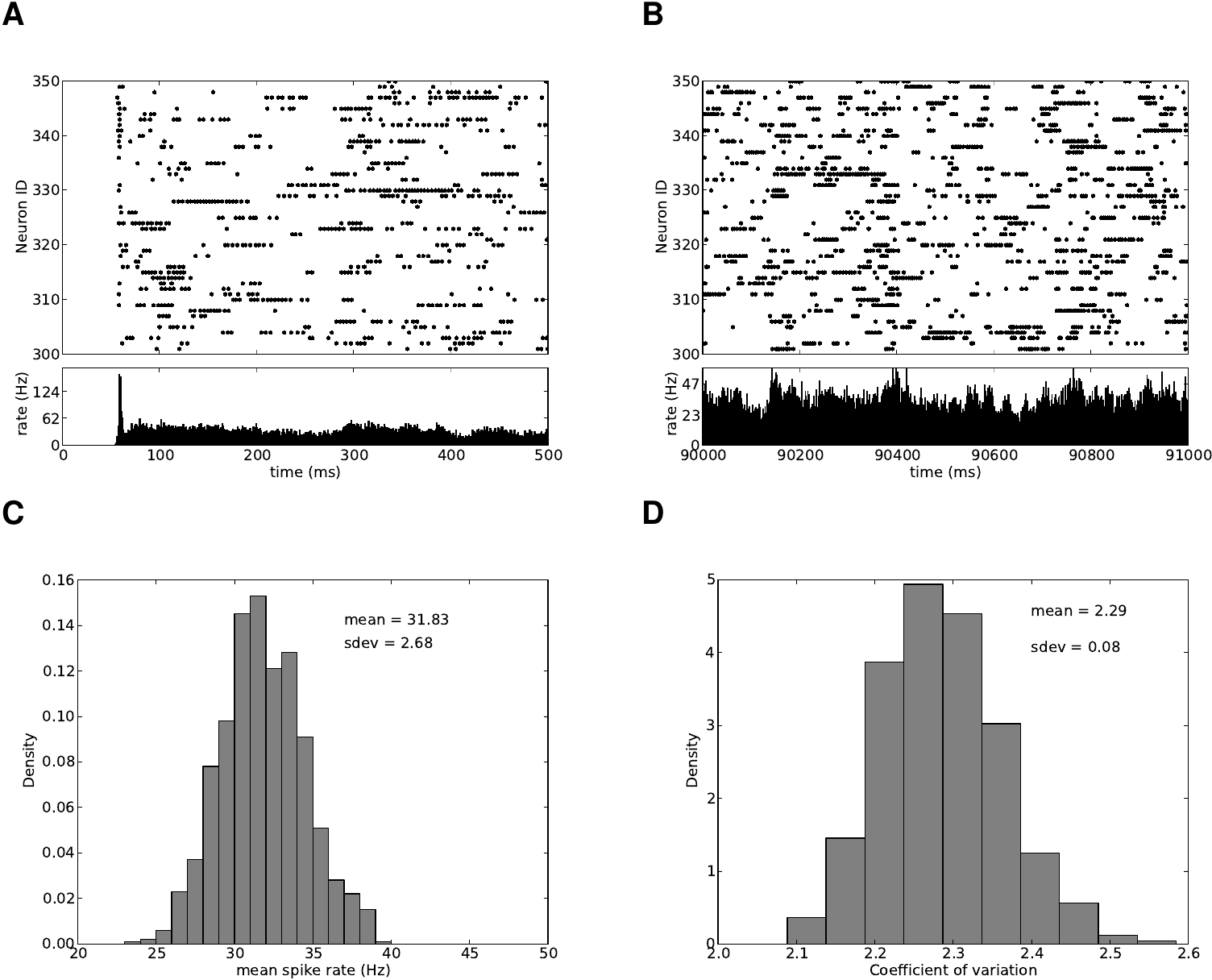}
\caption{\label{fig:self-sustained} Self-sustained activity in a
  network of $12,500$ neurons. Each panel shows the spiking activity
  of 50 representative excitatory neurons (top) and the population
  rate of $1\,000$ neurons (bottom). \panel{a} first $500\;ms$ of a
  simulation with parameters $g=5$, $J=4.0\;mV$, $\epsilon=0.01$ which
  lasted $100\;s$. Between $t=50\;ms$ and $t=200\;ms$, the neurons
  were supplied a weak Poisson stimulus to trigger spiking. After
  that, spiking activity continues for the whole simulation epoch at
  an average rate of $31.83\;$Hz. \panel{b} Activity of the same
  network as in A after $90\;s$. \panel{c} Distribution of the mean
  firing rates of the neurons. The mean firing rates follow a Gaussian
  distribution with mean $\mu=31.83\;$Hz and standard deviation
  $\sigma = 2.28\;$Hz. \panel{d} Distribution of the coefficients of
  variation (CV) of the interspike intervall (ISI) distribution. The
  CV follows a Gaussian distribution with mean $\mu_{CV}=2.29$ and
  standard deviation $\sigma_{CV}=0.08$. The rate and CV distributions
  were estimated from the activity of $1\,000$ neurons, recorded for 100
  seconds.}
\end{figure}

Figure \ref{fig:self-sustained} shows the results of a simulation that
lasted $100\;s$. 
Figure \ref{fig:self-sustained}A shows the first spiking activity of
50 excitatory neurons during the first $500\;ms$ (top) and the
corresponding ensemble rate of $1\,000$ excitatory neurons (bottom).
Again, at $t=0$, all neurons are in the quiescent state. To trigger
activity, we stimulated each neuron with weak Poisson noise for a
period of $t=150\;ms$. After that, no further input was given and the
network was left to its own devices.

The noise stimulus causes a transient network response, after which
the network settles to a mean rate firing rate of $32\;$Hz. This state
persists until the end of simulation at $t=100\;s$. Figure
\ref{fig:self-sustained}B shows the spiking activity after 90 seconds
of simulation to illustrate that the self-sustained state is stable.

There are a number of obvious differences between the raster plots of
the Brunel network in figure \ref{fig:brunel} and the raster plots in
figure \ref{fig:self-sustained}. 
First, the Brunel network is only active with external Poisson input,
whereas the network in figure \ref{fig:self-sustained} continues to fire
even in the absence of external input.  This state persists for very
long times, in this case 100 seconds. We call this network state the
\emph{self-sustained asynchronous irregular} (SSAI) state.
 
The self-sustained asynchronous irregular state is not stable in the
strict mathematical sense, because the random connectivity may cause
the number of active neurons to drop below a critical limit and then
spiking will stop altogether.  Since there is no noise input into the
network, it depends on the instantiation of the random connectivity if
and when the self-sustained state will stop (see also
\cite{Kumar2008,Destexhe2008}).  

Second, in figure \ref{fig:brunel} the spikes are relatively
homogeneously distributed, whereas in figure \ref{fig:self-sustained}
the distribution of spikes is very irregular.  By contrast, neurons in
the SSAI state exhibit large periods with few or no spikes at all, as
well as \emph{bursting} periods where spikes are so close together
that they can hardly be distinguished. It appears as if the neurons
switch between states of high activity and states of silence.

Third, the population rates (lower part of each panel) of the
self-sustained networks differ from the Brunel case. In Brunel's case
(figure \ref{fig:brunel}), the population rate stays quite constant
after the initial transient or it slightly oscillates
\cite{Kriener2008}. By contrast, the population rate of the
self-sustained networks in figure \ref{fig:self-sustained} show large
fluctuations without any obvious temporal structure.

To quantify the irregularity of neuronal activity we looked at the
distribution of mean firing rates (figure \ref{fig:self-sustained}C) as well as the
distribution of the coefficients of variation (figure \ref{fig:self-sustained}D). 

For figure \ref{fig:self-sustained}D, we computed the mean firing rates for $1\,000$ neurons over
the simulation period of 100 seconds. Firing rates are roughly
Gaussian distributed with a mean of $31.83\;$Hz and a standard
deviation $2.68\;$Hz. The lowest firing rate in the population is
$20\;$Hz and the highest rate is $40\;$Hz. 

From figures \ref{fig:self-sustained}A and \ref{fig:self-sustained}B
we know that over time, the firing rate of each neuron fluctuates
considerably. These fluctuations lead to a correspondingly large
coefficient of variation (CV), which is the standard deviation of the
interspike intervals (ISIs) divided by their mean. Figure
\ref{fig:self-sustained}D shows the distribution of the coefficients
of variation of $1\,000$ neurons. We find a Gaussian distribution with
mean $2.29$ and standard deviation $0.08$. The smallest CV is 2.1
which is more than twice as large as the CV of a Poisson process.

\subsubsection*{Interspike interval statistics in the SSAI state}

Next, we investigate the statistical properties of interspike
intervals in the SSAI state.  
If the neuron were firing Poissonian with rate $\lambda$ then their interspike intervals
should be exponentially distributed:

\begin{equation}
  f_{\exp}(t) = \lambda \exp\left(-\lambda \cdot t \right)
\end{equation}

On a logarithmic scale, the exponential distribution is a straight
line whose slope is proportional to the rate of the exponential
distribution, which is also the rate of the Poisson process:

\begin{align}
\log_{10}\left(f_{\exp}\left(t\right)\right) &= \log_{10}\left(\lambda \exp\left(-\lambda \cdot t \right) \right) \\
                               &= \log_{10}\left(\lambda\right) -
                               \lambda
                               \log_{10}\left(e\right)\cdot t\\
                               &= a\cdot t + b
\end{align}
with $a= -\lambda\,\log_{10}\left(e\right)$ and $b=\log_{10}\left(\lambda\right)$.

\begin{figure}[htp]
 \includegraphics{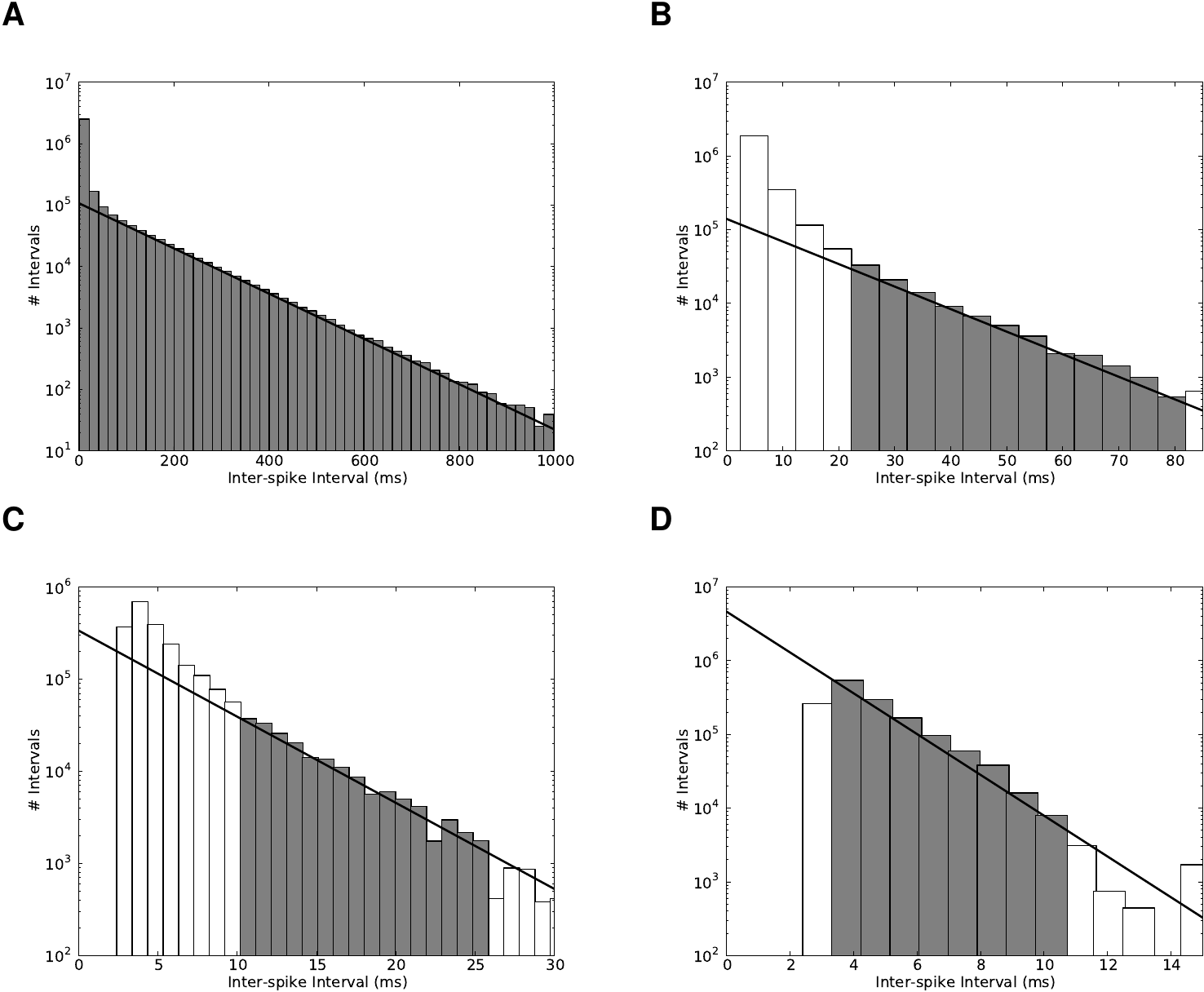}
 \caption{\label{fig:statistical-properties} Interspike interval
   distribution for the network in figure \ref{fig:self-sustained} on
   a logarithmic scale. In each panel, a regression line (black) shows
   which region of the distribution is well described by an
   exponential distribution. In panels B-D, only the gray bars were
   used to determine the regression lines. \panel{a} ISI distribution for
   all intervals smaller than $1\,000\;ms$. \panel{b} ISI distribution for
   intervals smaller than $80\;ms$, corrected by the intervals explained
   by the distribution in panel A. \panel{c} ISI distribution for
   intervals smaller than $30\;ms$, corrected by the intervals explained
   by the distributions in panels A and B.  \panel{d} ISI distribution
   for intervals smaller than $15\;ms$, corrected by the intervals
   explained by the distributions in panels A through C.  See text for
   details. }
\end{figure}

If the firing patterns in the SSAI state can at least partially be
described by a Poisson process, the logarithmic interspike interval
distribution should have a linear part.

Figure \ref{fig:statistical-properties}A shows the interspike interval
distribution for the network from figure \ref{fig:self-sustained} on a
logarithmic scale. For intervals larger than $80\;$ms, the distribution
is well described by a straight line with slope $a_{t>80}=-0.0037\;
ms^{-1}$, which corresponds to a firing rate of
$\lambda_{t>80}=8.81\,$Hz.

Intervalls smaller than $80\;$ms cannot be explained by this Poisson
process. In fact, almost 80\% (78.8\%) of all spikes contribute to
these intervals, although the neurons spend 80\% (79.2\%) of their
time in the remaining intervals larger than $80\;$ms.

To understand how this surplus of short intervalls is generated, we
return to the hypothesis that during SSAI activity the neurons are in
one of two states: a low rate state and a high rate state. 
The low rate state accounts for all intervalls larger than $80\;$ms,
while the high-rate state must account for the surplus of intervals
below the threshold.

Under this hypothesis, the interval distribution in figure
\ref{fig:statistical-properties}A should be the convolution of two
exponential distributions with rates $\lambda_1$ and $\lambda_2$. 
\begin{align}
  f_{ISI}(t) &= f_{\exp 1}(t) * f_{\exp 2}(t)\\
            &= \frac{\lambda_1 \cdot \lambda_2}{\lambda_1 - \lambda_2} \label{eq:isi_sum}
\left( \exp\left(-\lambda_2 t\right) - \exp\left(-\lambda_1 t\right)
\right)\\
\intertext{For $t \cdot \lambda_1 \gg 1$, we get}
f_{ISI}(t)   &\approx  \frac{\lambda_1 \cdot \lambda_2}{\lambda_1 - \lambda_2}
\left( \exp\left(-\lambda_2 t\right) \right) \label{eq:isi_approx}
\end{align}

Using equation \eqref{eq:isi_approx} we can estimate the slow firing
rate component. 

We can also estimate the fast firing rate component. To this end, we subtract the
expected number of intervals from the slow component, because from
equation \eqref{eq:isi_sum} we know that the interval distribution is
basically the sum of the two superimposed distributions. Thus:
\begin{align}
f_{ISI}(t) -  \frac{\lambda_1 \cdot \lambda_2}{\lambda_1 - \lambda_2}
\cdot f_{\exp 2}(t) &= \frac{\lambda_1 \cdot \lambda_2}{\lambda_1 -
  \lambda_2} \exp(-\lambda_1 t)
\end{align}

Figure \ref{fig:statistical-properties}B shows the ISI
distribution for intervals smaller than $80\;ms$ with the intervals,
generated by the slow rate subtracted. Again, a good part of the
distribution is well described by a Poisson process, this time with
rate $\lambda_{20<t<80}=68.20\,$Hz, but for intervals
shorter than $20\;ms$, the model again fails. 

We can repeat this procedure again to estimate the rate of the next faster
process. This is shown in figures \ref{fig:statistical-properties} C
and \ref{fig:statistical-properties}D.

Altogether we find, that the interval distribution of this network
can be described by four Poisson processes with widely different
rates, valid for different ranges of interspike intervals.

If the four Poisson processes were active in parallel, it would be
impossible to separate them in the interval distribution, because the
sum of two or more Poisson processes with rates $\lambda_1, \lambda_2,
\ldots, \lambda_n$ is again a Poisson process with rate
$\lambda_*=\sum_i \lambda_i$. Thus, we must assume that the neurons
switch between a small number of Poisson states, each with its own
firing rate and its own range of intervals.  These states and their
interval ranges are summarized in table \ref{tab:poisson_rates}.

The fastest process has a rate of $\lambda_1=786.41\;$Hz and is
responsible for the high-frequency burst with interspike intervals
between $3$ and $10\;ms$, observed in the raster plots of figure
\ref{fig:self-sustained}. Note that although this rate is larger than
the theoretical maximum of $\lambda_{\max}=1/t_{ref}=500.\;$Hz the
smallest interval is with $2.4\;ms$ still larger than the refractory
period of $2\;ms$. The very high rate of $\lambda_1$ is the result of the
large number of intervals between $3$ and $10\;ms$.

The slowest process is responsible for the large gaps between spikes,
observed in figure \ref{fig:self-sustained} and has a rate of
$\lambda_4=8.82\;$Hz.

\begin{table}
\begin{tabular}{lcrrrr}
\hline
No. & range & slope ($m$Hz) & rate ($$Hz) & $\sum$ time (\%) &
$\sum$ spikes (\%)\\
\hline\hline
1 & $3 < t \le 10$  & $-0.3415$ & $786.41$ & 4 & 28 \\
2 & $10 < t \le 25$ & $-0.1144$ & $263.48$ & 7 & 37 \\  
3 & $25 < t \le 80$ & $-0.0296$ & $68.20 $ & 6 & 11 \\
4 & $ 80 < t $      & $-0.0037$ & $8.82$   & 79& 21 \\
\hline
\end{tabular}
\caption{\label{tab:poisson_rates} Times scales and firing rates of
  a network during the SSAI state. The slopes correspond to the
  regression lines in figure \ref{fig:statistical-properties}.}
\end{table}

\subsubsection*{Firing rates and survival times depend on $J$ and $g$} 

So far we have looked at specific network configurations which were
suggested by the discrete Griffith model. We now step up from these
specific cases to see how the mean firing rate and the survival time
of the self-sustained activity states depend on the ration between
excitation and inhibition $g$ and on the synaptic strength $J$.

To this end, we simulated a network with $10\,000$ excitatory and 2500
inhibitory neurons for up to 100 seconds for combinations of
$(g,J)$, where $g$ was varied from $3$ to $6$ in steps of 0.01 and $J$
was varied from $1\;mV$ to $5\;mV$ in steps of $0.01\;mV$. In each simulation,
we measured the mean firing rate of the neurons as well as the
survival time of the SSAI state.

\begin{figure}[th]
 \includegraphics{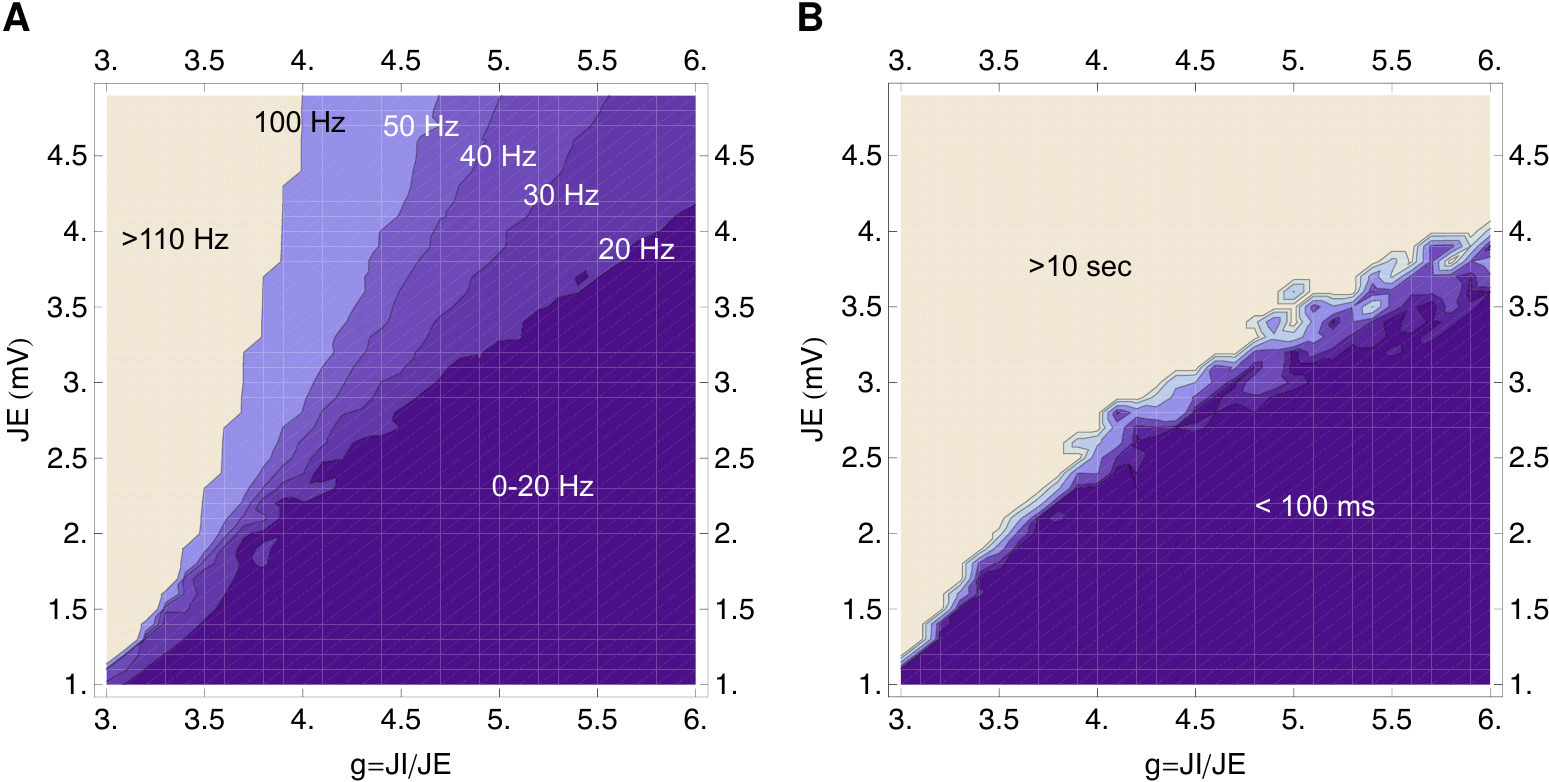}
 \caption{\label{fig:order4-contour} \panel{A} Contour plot of the mean firing rate as a
   function of $J$ and $g$. For small $g$ and $J>1$,
   the mean rate approaches the maximum $1/t_{ref}$. with increasing
   $g$, the mean rate decreases monotonically. The region between
   0--20 Hz is characterized by short-lived self-sustained
   states. Stable self-sustained states can be found for mean rates grater
   than $20\;$Hz. \panel{B} Contour plot of the survival time as a function of $J$ and
   $g$. There is a sharp transition between immediate
   network death (blue area) and long survival (beige area) which does
   not soften significantly with increasing $J$ and $g$.}
\end{figure}

The results of these simulations are summarized in figure
\ref{fig:order4-contour}. Figure \ref{fig:order4-contour}A shows the
contour plot of the mean firing rate as a function of $J_{E}$ and
$g$. Darker colors correspond to lower rates. Along the contour lines,
the firing rate remains approximately constant (iso-rate boundaries).

The abcissa represents ratio of inhibition to excitation $g$. For
$g=4$, excitation and inhibition contribute equally to the synaptic
current. For $g<4$, excitation contributes more than inhibition,
and for $g>4$, inhibition contributes more than excitation.

For $g<3$ the excitatory population dominates the network activity and
the firing rates approach the theoretical maximum of
$\nu_{max}=1/t_{ref}$. This is true even for small excitatory
amplitudes.

As $g$ increases with the inhibitory amplitude, rates become lower,
until they reach the physiologically interesting range of
0--40 Hz. This range is reached even before inhibition balances
excitation, however this activity is usually unstable. This is also
evident from the very steep transition between the white area of high
rates and the dark area of low rates.

With further increasing relative inhibition (larger $g$), the slope of
the transition between high and low rates decreases, creating wide
regions of intermediate rates.

Figure \ref{fig:order4-contour}B shows the survival time of network
activity for the same parameter range. Here, we observe a sharp
transition between basically two states: either activity ceases after
less than $100\;ms$, or it survives for a very long time. In contrast
to the firing rates in figure \ref{fig:order4-contour}A, the
transition remains sharp for all values of the relative inhibition
$g$. This means that SSAI states in inhibition dominated networks
($g>4$) are more stable than in balanced ($g=4$) or excitation
dominated networks ($g<4$).

\subsubsection*{Self-sustained states exist in a wide range of network sizes}

Self-sustained asynchronous irregular states are an emergent network
property which require a minimal network size. Thus, we were
interested in the smallest network which still show SSAI states. We
were also interested in larger networks where the number of connections
per neuron increases and the synaptic strength becomes smaller relative
to the firing threshold.

A trivial way of scaling the network is to start with a given
configuration (e.g. $N=10\,000$, $g=5$, and $J=3.8\;mV$) and to
increase the number of neurons, while keeping the synaptic amplitude
$J$ and the numbers of connections $C_E$ and $C_I$ constant. This
strategy works well, and yields qualitatively the same results as the
original smaller network (not shown). In particular the mean
firing rates stay constant.

Alternatively, we can keep the connection probability
constant. Then, the number of connections per neuron will grow as the
network increases. To compensate for the increasing number of
connections, we must decrease the synaptic amplitudes $J$ 
proportional to $\sqrt{N/N'}$. This scaling works until the synaptic
amplitudes become small compared to the threshold.

Similarly, there is a lower limit to the network size. If we reduce
the number of neurons for a fixed amplitude $J$ and fixed numbers of
connections $C_E$ and $C_I$, the connectivity $\epsilon$ increases
accordingly. Thus, we quickly reach the case where $\epsilon$
approaches $1$ and each neuron receives input from all other
neurons. In this state of total symmetry asynchronous irregular states
cannot exist.

\begin{figure}[t]
 \includegraphics{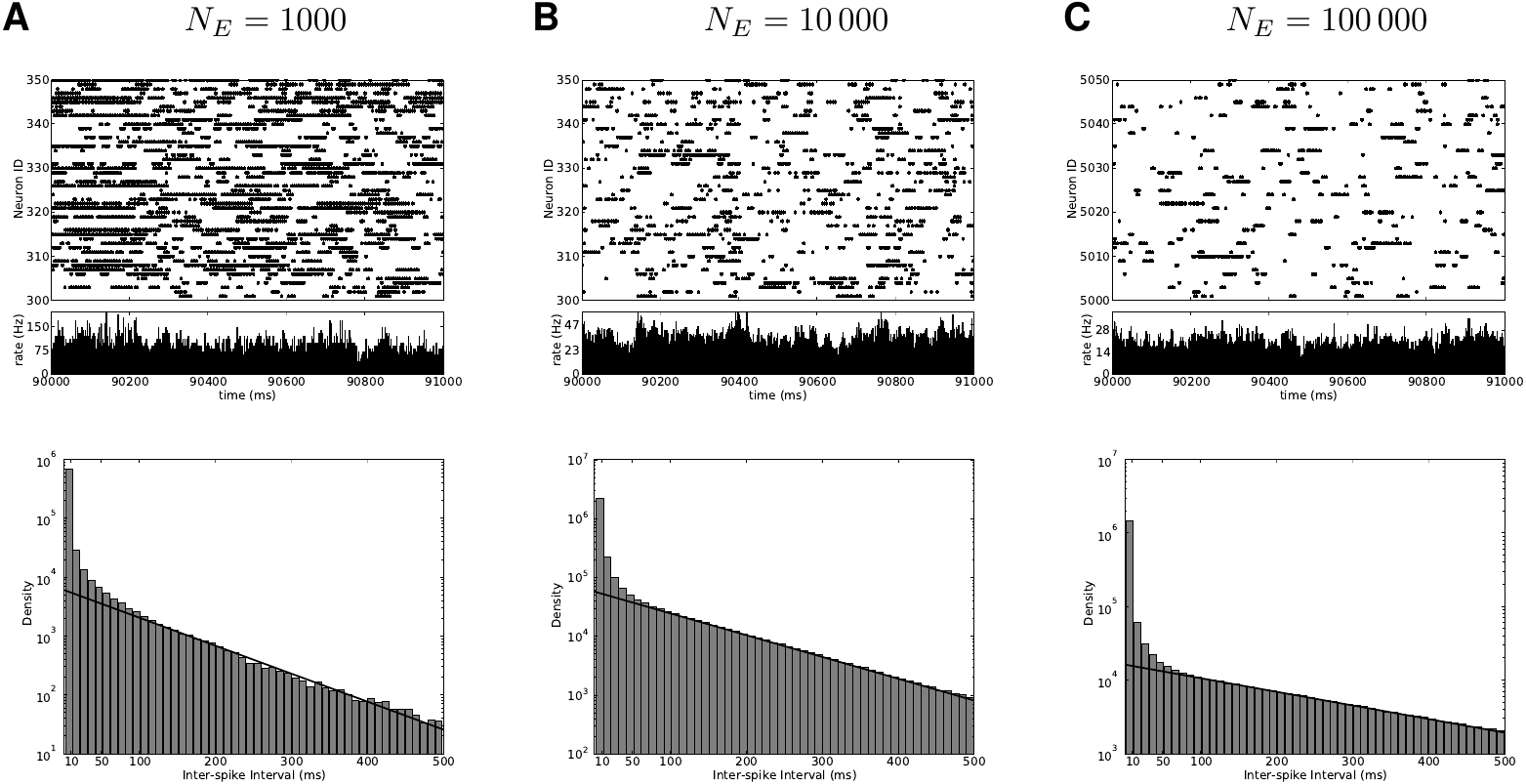}
 \caption{\label{fig:network-scaling} Self-Sustained activity in
   networks of different sizes. Raster plots of the spiking activity
   is shown in the top row, the interval distribution in the bottom
   row. The firing rates decrease with increasing network size. This
   is also visible from the interval distributions, that shift towards
   larger intervals. \panel{a} $1\,000$ excitatory and $250$
   inhibitory neurons, \panel{b} $10\,000$ excitatory and $2\,500$
   inhibitory neurons, and \panel{c} $100\,000$ excitatory and
   $25\,000$ inhibitory neurons.}
\end{figure}

If we keep the connectivity $\epsilon$ constant, we can, in theory,
reduce the network size until $\lfloor \epsilon*N\rfloor = 0$. Thus,
for $\epsilon=0.1$ the smallest theoretical network size would be
$N=10$. However, simulations show that already for $\epsilon=0.1$ and 
$N=100$ the self-sustained state requires synaptic amplitudes of
$J>\theta$ and shows unphysiological spike trains where each neuron
switches between high-frequency bursts and silence. Moreover, the
self-sustained state becomes unstable. The main reason seems to be
that for such small networks each neuron receives less than 10 inputs,
each sufficiently strong to trigger a spike. At the same time, each
neuron sees a considerable portion of the entire network. Thus,
correlations amplify quickly until coherent down-states stop the
network activity.
In between these extremes there is a wide range of network sizes and
connectivities where self-sustained states of asynchronous irregular
activity exist. 

Figure \ref{fig:network-scaling} shows the raster plots and interval
distributions for example networks of three sizes: $1\,000$, $10\,000$, and
$100\,000$ excitatory neurons.  The network in figure
\ref{fig:network-scaling}A has the highest rate and the largest PSP
amplitude, but the smallest number of connections per neuron.

The networks in figures \ref{fig:network-scaling}B and
\ref{fig:network-scaling}C have the same synaptic amplitudes and
the same connection probability, thus, they differ only in the number of
connections per neuron and in the resulting firing rates. 
Apart from the firing rates, their is no qualitative difference in the
raster plots or the interval distributions. The most prominent
feature seems to be that small networks are more dominated by small
spike intervals than large networks. This is best seen in the interval
distribution which widens considerably as the networks become larger.

\subsubsection*{Few strong synapses in a sea of weak ones}
\label{sec:DopedNetwork}
\begin{figure}[t]
 \includegraphics{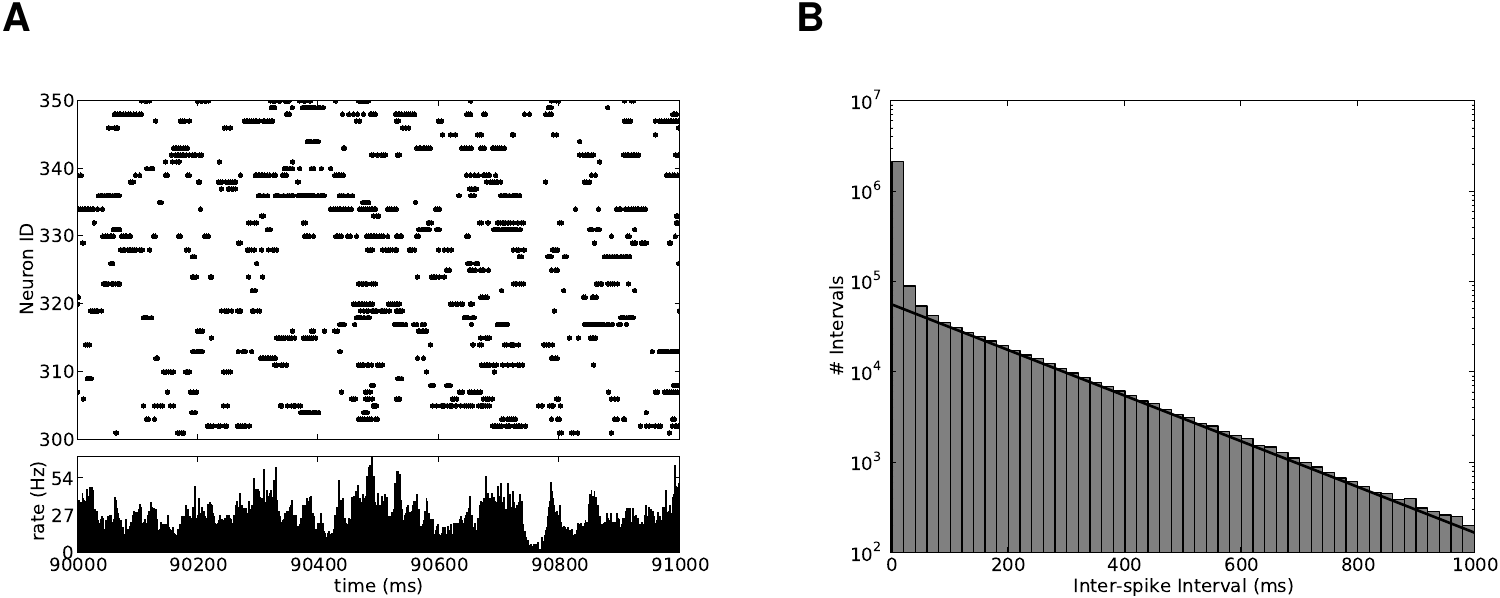}
 \caption{\label{fig:weak_and_strong} Self-Sustained activity in a network
   with many weak and few strong synapses. \panel{A} Raster plot of
   spiking activity (top) and population firing rate
   (bottom). \panel{B} ISI-histogram on a logarithmic scale.}
\end{figure}

Recently, it has been proposed that cortical networks have a long
tailed distribution of synaptic weights, where most connections are
very weak, but some connections are very strong  \cite{Song2005,Teramae2012}.
While the strong connections in such network might facilitate
self-sustained activity, the larger number of weak connections may
destroy these states again.
In the following, we tested, whether a netwok with many weak and a few
strong synapses will exhibit self-sustained activity states. To do so,
we took the Brunel network of figure \ref{fig:brunel} and increased
the weight of 1\% of its connections to $J=4\;mV$. 
The spiking activity of this network is shown in figure
\ref{fig:weak_and_strong}A and its ISI distribution in figure
\ref{fig:weak_and_strong}B. 

Figure \ref{fig:weak_and_strong}A shows the spiking activity after 90
seconds of simulation which means that a Brunel type network which is
\emph{doped} with a small number (1\%) of strong synapses also shows
self-sustained activity. The firing rate in this state is with 26 Hz
about 20\% lower than in the undoped network.

Figure \ref{fig:weak_and_strong}B shows the ISI-distribution on a
logarithmic scale, which clearly looks like the ISI distributions of
the plain SSAI networks. In the doped network it is also possible to
describe the ISI histogram as superposition of different states. Here,
the fastest state has a rate of $956\;$Hz which is assumed
for 3\% of the time. The slowest state has a rate of $5.85 $Hz and
accounts for 85\% of the time and almost 20\% of the spikes.

Obviously, even a small number of strong synapses suffices to
change the dynamics of the network. While the undoped
Brunel network, shown in figure \ref{fig:brunel} has a relatively
regular dynamics and cannot sustain firing without external input, the
doped network exhibits highly irregular activity which is autonomously
sustained without external input.
This dramatic effect of only a few strong synapses has important
implications for learning and plasticity.  On the one hand, it can be
beneficial, because it is easy to store information in self-sustained
states by facilitating just a tiny fraction of the synapses in a
sub-network. On the other hand, this effect can easily disrupt
processing, because otherwise silent populations can be easily pushed
into a self-sustained activity state.

\section*{Discussion}

In this paper, we show that self-sustained states of asynchronous
irregular activity can be induced in recurrent networks of simple
threshold elements, like integrate-and-fire neurons, under the
following conditions:
\begin{enumerate}
\item \label{condition1} a fraction of the connections are sufficiently
  strong (i.e. $\Theta/J \ll C_E$).
\item \label{condition2} a sufficient number of excitatory neurons is
  activated to trigger the self-sustained activity state.
\end{enumerate}

Condition \ref{condition1} contradicts the common view that the dense
cortical connectivity implies weak connection strengths
\cite{Abeles1991,Brunel2000}. Even if all connections in a recurrent
network are strong, the firing rate  will not approach the theoretical
limit of $1/\tau_{\text{ref}}$. Moreover, in a network with weak
synaptic connections, a small fraction of strong synapses will
actually decrease the overall firing rate in the network.

Condition \ref{condition2} implies that SSAI states can be induced by
temporarily lifting the network activity above the ignition point
(equation \eqref{eq:ignition-point}). This can be achieved by
supplying either a brief synchronized input to the excitatory neurons
or an asynchronous low-rate stimulus over a longer period of time.
Once the SSAI state is reached, the ignition stimulus may be switched
off and the AI state will persist.

\subsection*{Self-sustained activity as model for ongoing or
  spontaneous activity}

Spontaneous activity in the mammalian nervous systems is marked by low
firing rates between 5 and 10 Hz. By contrast, the self-sustained
activity states considered here show much higher rates of $20\;$Hz and
more. This raises the question if and how self-sustained
states give rise to low-rate spontaneous activity.

There are several possibilities and we will outline the two most
probable ones.

\paragraph{Leaking excitation} It has been repeatedly shown that
weakly coupled networks with random external input shows stable
low-rate firing. Thus, the simplest model to explain low rate
spontaneous activity, is excitation from a strongly coupled assembly
which acts as external drive for a weakly coupled network. 

\paragraph{Coupled assemblies} Consider a network of many
self-sustained networks (assemblies) which are coupled by competetive
inhibition such that at any point in time only one or a small fraction
of the assemblies can be active. During normal operation, each
assembly would be activated for a short time, before activity switches
to another assembly. As a result, the average firing rate of the
entire network will be much lower than within each of the assemblies.

\subsection*{Self-sustained activity in conductance based networks}
\begin{figure}[ht]
 \includegraphics{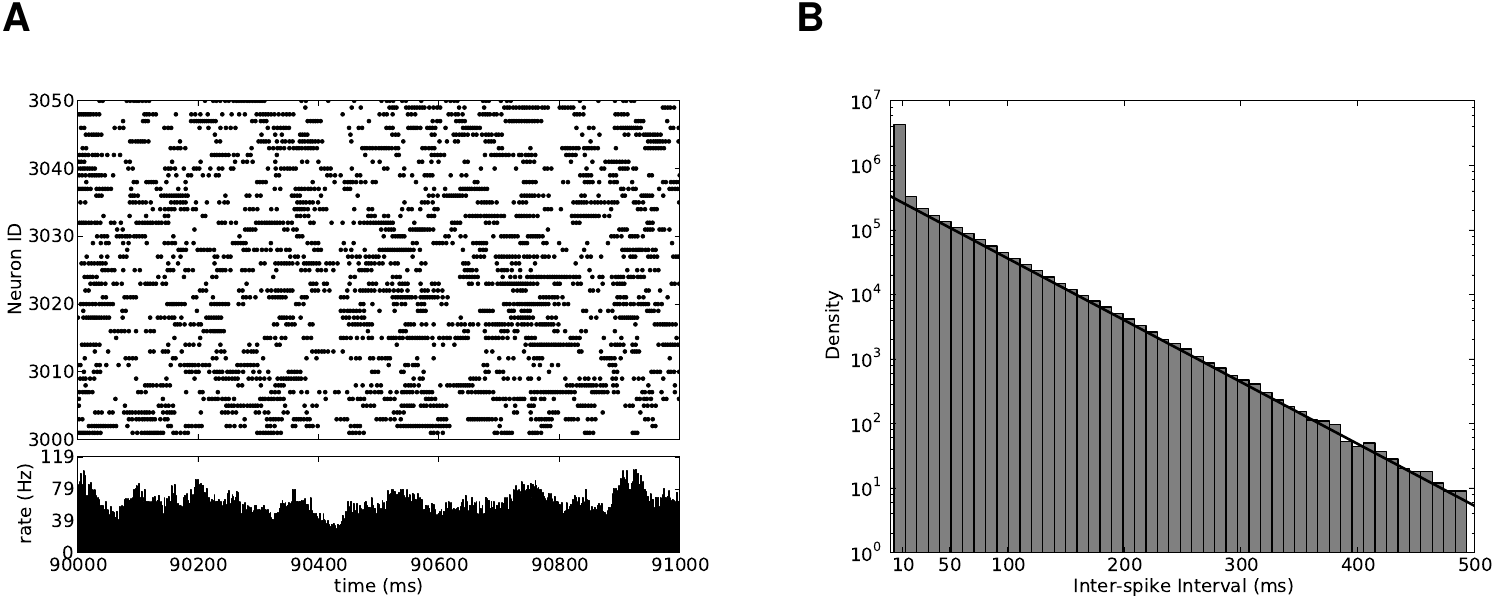}
 \caption{\label{fig:coba_net} Self-Sustained activity in a network
  of conductance based neurons.\panel{A} Raster plot of
   spiking activity (top) and population firing rate
   (bottom). \panel{B} ISI-histogram on a logarithmic scale.}
\end{figure}

The self-sustained activity states described here result from the
combinatorics of strong inputs. They do not rely on specific
properties of the synapses or the membrane dynamics. Thus, SSAI states
will occur under similar conditions in networks of more complex
conductance based neurons.  Kumar et al.\ reported self-sustained
activity in large networks of conductance based neurons. In their
model, they found that the survival time of self-sustained states
grows exponentially with the network size. For networks of $1\cdot
10^4$ neurons, they found a survival time of less than one second.

By applying the reasoning developed in this manuscript, we found that
also in networks of conductance based neurons with few strong
connections spiking activity persists for very long periods. Figure
\ref{fig:coba_net} shows the results of a simulation that lasted 100
seconds. Both raster plot (figure \ref{fig:coba_net}A) and ISI
distribution (figure \ref{fig:coba_net}B) look qualitatively identical
to that which we found in the current based case. This is a good
indication that the arguments and consequences of self-sustained
activity, induced by strong synaptic weights also apply to networks of
more realistic model neurons or even to networks in the brain.

\subsection*{Variability of firing}

It has been argued that that the high irregularity of cortical cells
is inconsistent with the temporal integration of random EPSPs
\cite{Softky1993a}, because such a \emph{counting process} would
inevitably be more regular than a Poisson process \cite{Shadlen1998}.
This phenomenon can be observed in Brunel's model in figure
\ref{fig:brunel} \cite{Barbieri2008}. The random superposition of
small synaptic weights leads to very regular firing patterns with
$CV_i \ll 1$.
Several authors have proposed amendmends to account for the high
irregularity of cortical firing \cite{Barbieri2007,Barak2007}, by
introducing additional mechanisms like different synaptic pathways,
synaptic fascilitation and depression, or spike-timing dependent
plasticity.  

In this paper, we demonstrated that none of these mechanisms is
actually needed to obtain highly irregular firing. It suffices to take
a sparsely connected random network of integrate-and-fire neurons with
a small fraction of strong connections. These lead to activity states
with $CV_i > 1$, more consistent with the irregularity of
experimentally observed ISI distributions \cite{Compte2003}.

We found that during SSAI activity, the neurons switch between several
discrete states which results in the observed high-variability of the ISIs.

\subsection*{Survival and survival time}

Kumar et al.\ \cite{Kumar2008} observed that the survival time of
self-sustained activity states in networks of conductance based
neurons grows exponentially with the network size. Networks of
$10\,000$ neurons could sustain firing for as little as 10 ms, while
larger networks of $50\,000$ neurons sustained firing for up to 10
seconds.

By contrast, we observed little dependence of the survival time on the
network size. Rather, the networks discussed here show bimodal
survival times: Either the activity ceases after a few milliseconds,
or it persists for many seconds or minutes (see figure
\ref{fig:order4-contour}B). This was also true for the conductance
based network of figure \ref{fig:coba_net} which means that the SSAI
state described here is different from the persistent activity
described by Kumar et al.\ \cite{Kumar2008}.

The strong coupling between neurons results in activity which is
determined by the combinatorics of the connectivity matrix (the
network). Since there is no external source of noise, the response of
a given network to a particular stimulus is deterministic. At each
point in time a different subset of the connections carries the
activity forward. Only if one of these subsets is too small to fulfill
the threshold criterion will the activity cease.

\subsection*{Burstiness and multi-state interspike interval distributions}

A common measure for burstiness of firing is the surplus of small ISIs
(e.g. ISIs $<$ 5 ms) compared to a Poisson process of the same rate
\cite{Kaneoke1996,Compte2003}. According to this criterion, the firing
patterns in SSAI networks is clearly bursty. Moreover, this burstiness
is a pure network phenomenon, since the underlying integrate-and-fire
neuron model is too simple to produce bursts \cite{Gerstner2002}.

We have also shown that bursts in SSAI networks are the signature of
different states, each characterised by a different Poisson rate. In
total, we observed up to 4 different states (figure
\ref{fig:statistical-properties}).

A similar decomposition of interspike interval distributions of
experimental data was described by Britvina and Eggermont
\cite{Britvina2007}, who showed that the interspike intervals of layer
II/III neurons from cat primary auditory cortex can be described by
a Markov model, switching between different disjunct states.

\subsection*{Self-sustained activity as memory}

Self-sustained activity states do not require very large networks, but
can be found in networks larger than about 100 neurons. Moreover,
self-sustained states are very stable and persist for many seconds.
Thus, a small population of a few hundred neurons could already store
information in its activity state, which is very economic compared to
other attractor memory models \cite{Roudi2007}.
However, in the presence of strong connections, even small amounts of
activity \emph{leaking} into the memory population will trigger the
SSAI state, irrespective of a stimulus. This poses a serious problem
for models that store information in the activity state of a neural
population. To be reliable, such memories need an additional control
mechanism which prevents spurious activation of memory populations by
the embedding network. Such a control mechanism could be provided by
inhibition between different memory pools so that at any time only one
or a few memories can be active \cite{Deco2005,Renart2007}.

\subsection*{Self-sustained activity as signal propagation}

In self-sustained networks, the only source of randomness is the
connectivity matrix. Once the connectivity is fixed, the activity in
the network is deterministic.
Consequently, we may describe self-sustained activity as a sequence of
activated neuron groups: An initially activated group of neurons $G_0$
will trigger spikes in another group of neurons $G_2$ which in turn
will activate the group $G_3$ and so on.

Since the connections are strong, activating $G_0$ will always result
is the same activation sequence 
\[ 
 G_0\rightarrow G_1 \rightarrow \ldots \rightarrow G_N
\]
and a different starting group $G'_0$ will result in a
different activation sequence
\[ 
 G'_0\rightarrow G'_1 \rightarrow \ldots \rightarrow G'_N.
\]
Thus, we may regard  the initially activated neuron group $G_0$ as a
\emph{signal} which propagates through the network.

An alternative interpretation of the self-sustained activity is that
of \emph{recalling} a sequence which has previously been stored by an
appropriate learning mechanisms. 

In this respect, strongly coupled networks differ from other network
architectures like weakly coupled
networks and synfire chains \cite{Abeles1991}. 

In weakly coupled networks, signal propagation is difficult due to the
weak connections and the high level of noise \cite{Vogels2005}.

Strongly coupled networks could also be interpreted as an intertwined
synfire chain, because both are characterized by a fixed sequence in
which neurons are activated. However, there is an important
difference: Activity in strongly coupled networks is not only
deterministic, it is also chaotic \cite{Vreeswijk1996}. Thus, changing
only one neuron in $G_0$ will quickly result in an activation sequence
$G'_1 \rightarrow G'_2\rightarrow \ldots$ which is distinct from the
original activation sequence. By contrast, synfire chains are robust
to small variations in their activation sequence
\cite{Abeles1991,Diesmann1999}. If one or a few neurons are missing or
replaced in the initial set $G_0$, the divergent/convergent
architecture of the synfire chain will \emph{repair} this defect in
the subsequent activation stages \cite{Gewaltig2001c}.

\section*{Methods}
\label{sec:methods}

\paragraph{Neuron and network parameters}
\begin{figure}[htp]
 \includegraphics[width=0.9\textwidth]{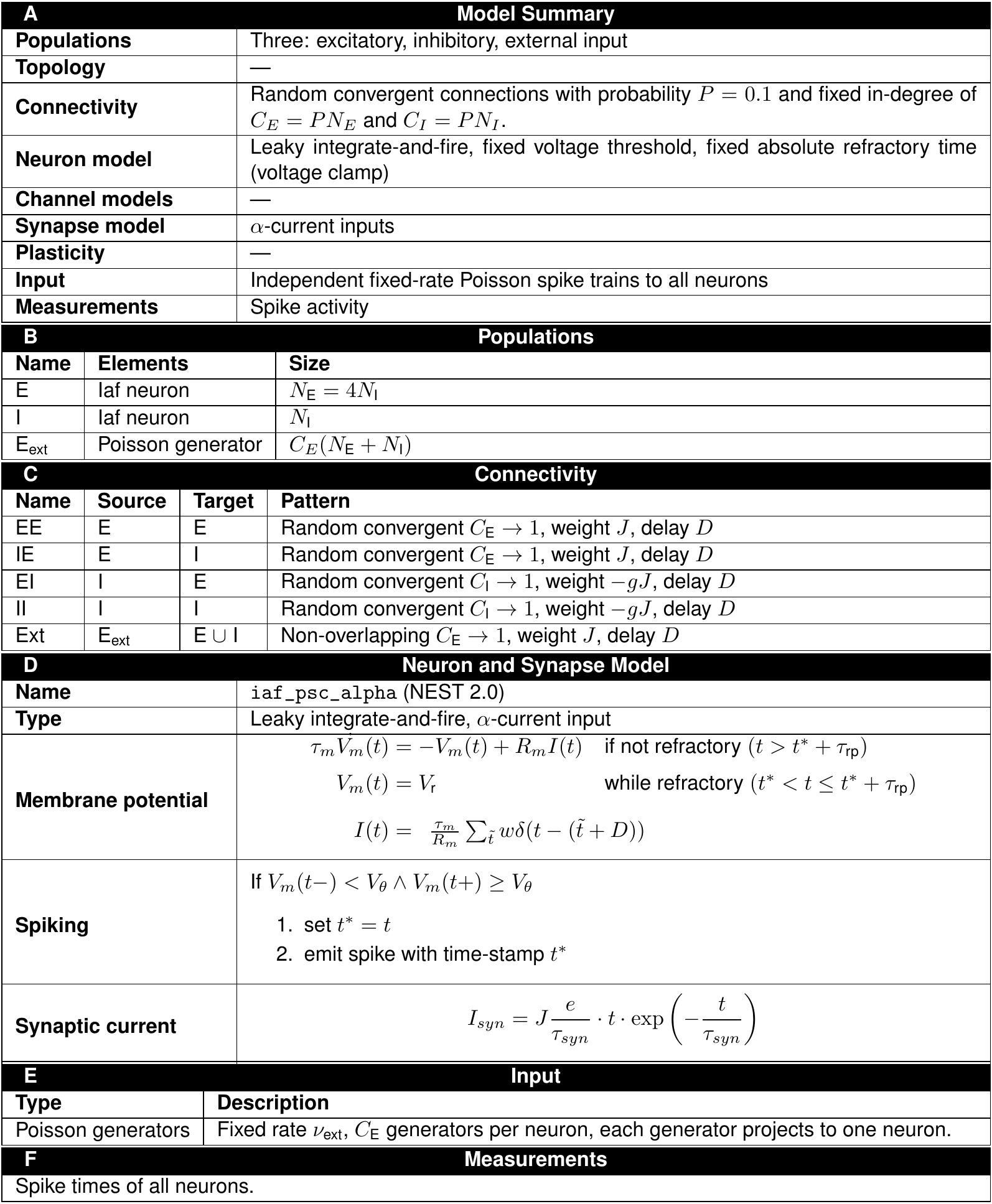}
 \caption{\label{fig:netparams} Model description according to \cite{Nordlie2009}, part 1.}
\end{figure}
\begin{figure}[htp]
 \includegraphics[width=0.9\textwidth]{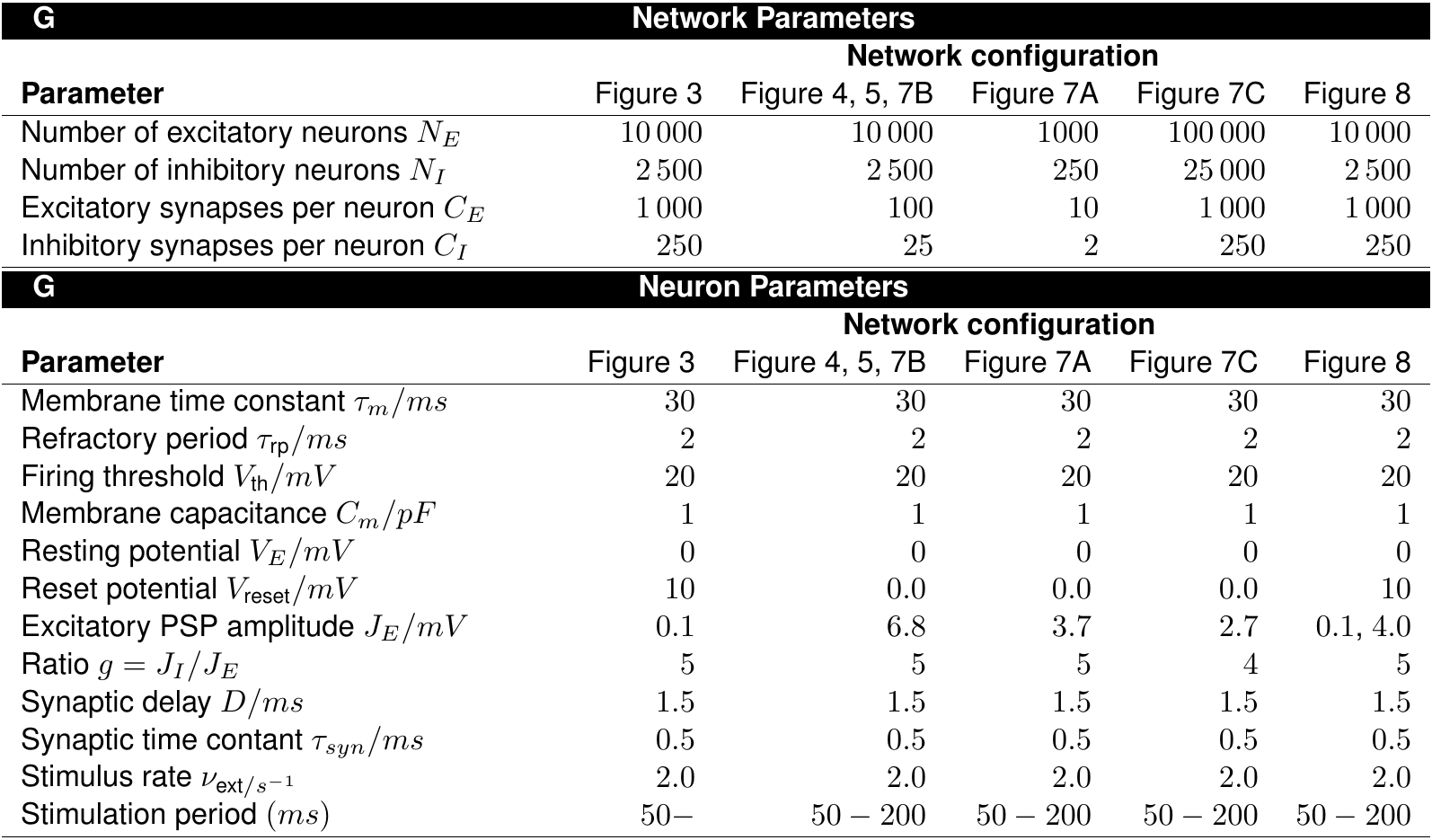}
 \caption{\label{fig:neuronparams} Model description according to \cite{Nordlie2009}, part 2.}
\end{figure}

All networks were simulated using a current based integrate-and-fire
model where synaptic currents are modeled as alpha functions. The
neuron model and network architecture are described in figures
\ref{fig:netparams} and \ref{fig:neuronparams} according to the
recommendations of Nordlie et al. \cite{Nordlie2009}.  

Brunel \cite{Brunel2000} originally used an integrate-and-fire model
with delta currents. These lead to instantaneous voltage jumps in
response to each spike. In networks with weak synaptic couplings, such
as Brunel's network, these discontinuities are smeared out by the
noise. In networks with strong synaptic couplings, however, the
voltage jumps remain visible in the interval and rate distributions,
since most spikes are then bound to the time grid imposed by the
synaptic delays. Thus, we used the more realistic alpha-functions as
synaptic currents.

\paragraph{Data acquisition}
Spike data was recorded to file from the first $1\,000$ non-input neurons
and analyzed offline. Since the network was connected as a random
graph, we could record from a consecutive range of neurons without
introducing a measurement bias.

\paragraph{Spike data}
For each neuron $i$, we record the sequence of spike times
$\mathsf{S}_i=\left\{ t_1, t_2, \ldots \right\}_i$ over an observation
interval $T$. Unless stated otherwise, the observation interval was
$100\;s$.  In the following, we write spike-trains as
\begin{equation}
s_i(t)=\sum_{t' \in \mathsf{S}_i} \delta(t-t')
\end{equation}

\paragraph{Firing rates}
The average firing rate of a neuron $i$ was determined by dividing the
total number of spikes $n_i$ within the observation interval $T$ by the its
duration:
\begin{align}
n_i  &= \int_T s_i(t') dt'\\
< \nu_i > &= \frac{n_i}{T} \label{eq:firing-rate}
\end{align}

Since there is no noise disturbing the neurons, each independent
simulation run (trial) yields the same spike trains. 
The random connectivity, however, allows us to interpret different
neurons as independent realizations of the same random process. Thus,
the instantaneous population rate is a good approximation of the
average instantaneous firing rate of a neuron.

We computed the population rate, by summing the spikes of all observed
neurons within a $\Delta T=1\;ms$ window around the time of interest:
\begin{equation}
\nu_N(t,\Delta t)= \frac{1}{N}\sum_{i=1}^N \int_t^{t+\Delta t} s_i(t')dt'
\end{equation}
The refractory period of $1\; ms$ ensures that each neuron contributes
at most one spike to the population rate.

\paragraph{Rate distribution}
To compute the distribution of firing rates, we first compute the
average firing-rate of each neuron according to \ref{eq:firing-rate}
and then construct a histogram with bin-size $\Delta r=1\;$Hz from the
set of all firing-rates:
\begin{equation}
H_{\nu}(r,\Delta r) = \frac{1}{N}\sum_{i=1}^N \int_r^{r+\Delta r} \delta(\nu_i-r')\;
dr'
\end{equation} 

\paragraph{ISI distribution}
Given the ordered sequence of spike times $\mathsf{S}_i$ of neuron
$i$, we construct the set of interspike intervals as:
\begin{equation}
\mathsf{ISI}_i=\{t_2-t_1, t_3-t_2,\ldots\}_i={\tau_1, \tau_2,\ldots,\tau_{\#(ISI_i)}}_i
\end{equation}

The ISI distribution was then constructed by counting the number of
intervals that fall in consecutive $1\;ms$ bins in a histogram. To
improve the statistics of the histogram, we combined the intervals of
all recorded neurons.
\begin{align}
H_{ISI,i}(isi, \Delta isi)&= \frac{1}{\#(\mathsf{ISI}_i)} \sum_{\tau \in \mathsf{ISI}_i}
\int_{isi}^{isi+\Delta isi} \delta(\tau-\tau')\; d\tau'\\
H_{ISI}(isi,\Delta isi) &= \frac{1}{N}\sum_{i=1}^N H_{ISI,i}(isi, \Delta isi)
\end{align} 

\paragraph{CV distribution}
To compute the CV distribution, we first computed the ISI
distributions for each neuron $i$ individually, according to the
procedure described above. We then determined the coefficients of
variation for each neuron $i$ according to:
\begin{equation}
  CV_i\frac{\sqrt{<ISI_i^2>-<ISI_i>^2}}{<ISI_i>}
\end{equation}

\paragraph{Simulation and analysis}
All simulations were done with the Neural Simulation Tool NEST
\cite{Gewaltig2007}, using its Python interface pyNEST
\cite{Eppler2008a}. The simulation data was written to disk and
analyzed off-line, using the NumPy and SciPy libraries for
Python (see \texttt{http://www.scipy.org} and \texttt{http://www.python.org}).
 
\section*{Acknowledgements}
The author would like to thank Gaute Einevoll, H{\aa}kon Enger, Morits
Helias, Edgar K\"orner, Birgit Kriener, Arvind Kumar, Hans-Ekkehard
Plesser, and Tom Tetzlaf for discussions and comments.
\bibliography{SelfSustained}
\end{document}